\documentclass[12pt, a4paper]{article}
\usepackage[english]{babel}

\usepackage[a4paper,top=2cm,bottom=2.5cm,left=2.5cm,right=2.5cm,marginparwidth=1.75cm]{geometry}
\usepackage{amsmath,amsfonts}
\usepackage{caption,adjustbox}
\usepackage{subcaption}
\usepackage{graphicx}
\usepackage[colorlinks=true,linkcolor=NavyBlue,citecolor=NavyBlue,urlcolor=Blue]{hyperref}
\usepackage[usenames,dvipsnames]{xcolor}
\usepackage{physics}
\usepackage{tikz}
\usepackage[affil-sl]{authblk}

\usepackage[symbol]{footmisc}
\usepackage{comment}
\usepackage{csquotes}
\usepackage{hyperref}
\usepackage[normalem]{ulem}
\usepackage{fancyhdr}

\usepackage[style=numeric-comp,sorting=none,maxbibnames=9]{biblatex}
\title{
\textbf{\textcolor{Maroon}{\Large The Effective Theory of Quantum Black Holes}}}
\author[1]{\small E. \textsc{Binetti}\thanks{\href{mailto:e.binetti@studenti.unina.it}{e.binetti@studenti.unina.it}}}
\author[2,3]{\small M. \textsc{Del Piano}\thanks{\href{mailto:manuel.delpiano-ssm@unina.it}{manuel.delpiano-ssm@unina.it}}}
\author[4]{\small S. \textsc{Hohenegger}\thanks{\href{mailto:s.hohenegger@ipnl.in2p3.fr}{s.hohenegger@ipnl.in2p3.fr}}} 
\author[3]{\small F. \textsc{Pezzella}\thanks{\href{mailto:franco.pezzella@na.infn.it}{franco.pezzella@na.infn.it}}} 
\author[1,2,3,5,6]{{\small F. \textsc{Sannino}}\thanks{\href{mailto:sannino@cp3.sdu.dk}{sannino@cp3.sdu.dk}}}

\affil[1]{\small Dept. of Physics E. Pancini, Universita di Napoli Federico II, via Cintia, 80126 Napoli, Italy}
\affil[2]{\small Scuola Superiore Meridionale, Largo S. Marcellino, 10, 80138 Napoli NA, Italy}
\affil[3]{\small INFN sezione di Napoli, via Cintia, 80126 Napoli, Italy}
\affil[4]{\small Institut de Physique des 2 Infinis (IP2I), CNRS/IN2P3, UMR5822, 69622 Villeurbanne, France Universit\'e de Lyon, Universit\'e Claude Bernard Lyon 1, 69001 Lyon, France}
\affil[5]{\small $CP^3$-Origins \& D-IAS, Southern Denmark Univ., Campusvej 55, 5230 Odense M, Denmark}
\affil[6]{\small CERN, Theoretical Physics Department, 1211 Geneva 23, Switzerland}
\date{}

\let\oldheadrule\headrule
\renewcommand{\headrule}{\color{white}\oldheadrule}
\begin{document}
\maketitle

\thispagestyle{fancy}
\rhead{CERN-TH-2022-048}

\begin{abstract}
We explore the quantum nature of black holes by introducing an effective framework that takes into account deviations from the classical results. The approach is based on introducing quantum corrections to the classical Schwarzschild geometry in a way that is consistent with the physical scales of the black hole and its classical symmetries. This is achieved by organizing the quantum corrections in inverse powers of a physical distance. By solving the system in a self-consistent way we show that the derived physical quantities, such as event horizons, temperature and entropy can be expressed in a well defined expansion in the inverse powers of the black hole mass. The approach captures the general form of the quantum corrections to black hole physics without requiring to commit to a specific model of quantum gravity. 
\end{abstract}

\maketitle

\section{Introduction}

Understanding the quantum nature of space-time is an open challenge both from a theoretical and an  experimental point of view. Quantum gravity effects are thought to be relevant, for example, in gravitational collapse of astrophysical objects as well as evaporation processes of Planck-size black holes (BH). The goal of this work is to construct an effective framework that allows to investigate quantum corrections for BH physics in order to extract reliable predictions. Effective approaches have been applied extensively to account for quantum corrections in gravity and particle physics, see \cite{Burgess:2003jk} for an overview. 
 
Rather than considering a specific theory of quantum gravity, our philosophy is to develop a general effective framework based on formulating BH metrics via dimensionless quantities and their physical scalings (see \cite{ChenGoldenfeldOono1995,PhysRevE.49.4502,Barenblatt} for related ideas in different areas of physics). Although this approach can be viewed as a renormalization improvement \cite{Coleman:1973jx,Migdal:1973si,Gross:1973ju,Pagels:1978dd,Matinyan:1976mp,Adler:1982jr,Dittrich:1985yb} of the BH metrics, it differs from the Wilsonian interpretation \cite{Wilson:1971bg,Wilson:1971dh} of the running of couplings of the effective action \cite{Bonanno:2000ep} further explored in \cite{Platania:2019kyx}, and therefore it is compatible with the arguments of Ref.~\cite{Donoghue:2019clr}.

We elucidate  our approach by focusing  on the static spherically symmetric classical Schwarzschild  BH. {After introducing the approach}, we  determine the impact of the leading order quantum corrections on the physical quantities such as event horizons, temperature and entropy in a consistent fashion. In the way the framework is setup, quantum corrections to physical observables appear as a well-defined expansion in  the mass of the BH relative to the classical results. We show that the approach can be consistently generalized to higher order quantum corrections leading to higher mass suppressed corrections. Although we do not discuss it in this work,  our approach can be further generalized to account for non-local corrections to effective gravity actions  \cite{Donoghue:1993eb,Donoghue:1994dn,Donoghue:1995cz,Donoghue:2015xla,Donoghue:2014yha,Donoghue:2015nba,Barvinsky:1985an,Barvinsky:1983vpp,Barvinsky:1987uw,Barvinsky:1990up,Barvinsky:1994ic,Avramidi:1990je,Codello:2014sua,El-Menoufi:2015cqw,Capozziello:2021bki,Capozziello:2022lic}. Our findings amount to establishing a self-consistent effective counting scheme, based on the physical mass of the BH. 

The work is organized as follows. In Section~\ref{sec. schwarzschild} we start by introducing the effective framework and by setting up the notation. The Section is further divided into several subsections. The rationale behind our way to upgrade the classical metric to an effective quantum one is summarized in the first subsection.  We then move to determine the leading order quantum corrected horizons and discuss their impact on the BH physics. We show that, depending on the sign of the first leading order corrections, the geometry can develop a second (internal) horizon.  
We then move to show the associated conformal diagrams.  The self-consistency of the approach, when considering the backreaction stemming from the quantum corrected proper distance, is presented in ~\ref{QGD}. In Section~\ref{HOC} we discuss how to take into account higher order quantum gravity corrections to the metric. One of the main results is establishing an effective consistent framework in computing quantum corrections to BH physics organised in their mass expansion. 
 
The quantum corrected thermodynamic properties, such as temperature and entropy, are discussed in Section~\ref{Thermo}. Conclusions and outlook are offered in Section~\ref{Conclusions} while in the Appendix~\ref{Appendix} we provide further details on how to compute the horizons in our framework. 

\section{Quantum Schwarzschild Black Hole} \label{sec. schwarzschild}
We focus on the simplest BH {in four dimensions,} featuring a spherical and stationary geometry with Schwarzschild metric 
\begin{equation}\label{metrics}
    \dd s^2=-f_0(r) \dd t^2 + \frac{\dd r^2}{f_0(r)}+r^2 \dd \theta^2 + r^2 \sin^2 \theta \, \dd \phi^2\, ,
\end{equation}
where we use spherical coordinates and the metric tensor depends only on the radial one through the function 
\begin{equation} \label{metricc}
    f_0(r)=1-\frac{2 G_\mathrm{N} M}{r}\, ,
\end{equation}
with  $M$ being the mass sourcing the gravitational field and $G_\mathrm{N}$ the Newton constant. 

\subsection{Quantum Framework} 
\label{quantum}
We now upgrade the metric (\ref{metrics}) to a quantum one without committing to a specific underlying quantum gravity theory\footnote{At the quantum level, we require the existence of a spherically symmetric metric with a time-like Killing vector. This ensures that the quantum metric still preserves the form \eqref{metrics}
\begin{equation}\label{metrics footnote}
    \dd s^2=-h(r) \dd t^2 + \frac{\dd r^2}{f(r)}+r^2 \dd \theta^2 + r^2 \sin^2 \theta \, \dd \phi^2\, .
\end{equation}
Here $f$ and $h$ are two functions of $r$. In this work, since we compute static properties we focus on the quantum corrections contained in $f(r)$.}. At the classical level the metric depends on two dimensionful quantities\footnote{The Planck scale hidden in the Newton constant at the classical level defines the units, and thereby has no influence on the classical physics.}, the mass of the BH and the coordinate $r$. In the following we describe  the quantum framework that we employ to determine quantum corrections to BH observables. 
\begin{enumerate}
  \item The quantum corrections are controlled by the Planck length $\ell_{\rm P}= 1/M_\mathrm{P}$ (with $M_\mathrm{P}$ the Planck mass), which governs the transition from the classical to the quantum regime. As such, now $\ell_{\rm P}$ is upgraded to a physically relevant length beyond providing just a unit of measure. To reflect this, we introduce the following dimensionless quantities: \begin{equation}\label{eq: dimless z}
    z:=M_\mathrm{P}r=\frac{r}{\ell_\mathrm{P}}\,  ,\qquad \chi:=\frac{M}{M_\mathrm{P}}\, ,
    \end{equation} and rewrite \eqref{metricc} as: 
    \begin{equation} \label{metricz}
    f_0(z) = 1 - \frac{2\chi}{z} g \ , \qquad {\rm with} \qquad g := G_\mathrm{N}  M_\mathrm{P}^2 = 1 \ .
    \end{equation}
    \item {Transitioning from the classical to the quantum regime} requires to modify \eqref{metricz} as follows: 
    \begin{equation} \label{lapse-function}
        f(z,\frac{u}{\ell_\mathrm{P}}) = 1 - \frac{2\chi}{z} g (z,\frac{u}{\ell_\mathrm{P}}) \ ,
    \end{equation} 
    where $g$ is an a priori undetermined function.\footnote{As already in the classical case (\ref{metricc}), we do not explicitly exhibit the dependence on $\chi$} Here $u$ is an arbitrary renormalization scale required to compensate for the presence of a fundamental length in the problem, i.e. $\ell_\mathrm{P}$. Since $u$ is arbitrary no physical quantity can depend on it. This means that the derivative of any such quantity with respect to $u$ must vanish, therefore imposing  non-trivial consistency conditions also on $\displaystyle{g (z,\frac{u}{\ell_\mathrm{P}})}$ (see \emph{e.g.} \cite{ChenGoldenfeldOono1995} for similar arguments in other physical systems). In order for any allowed coordinate transformation of $f_0(z)$ to be carried over at the quantum level one has to conclude that $g$ is a protected quantity and therefore: 
    \begin{equation}
        g(z,\frac{u}{\ell_\mathrm{P}}) \quad \longrightarrow  \quad   g(d,\frac{u}{\ell_\mathrm{P}})  \ ,
    \end{equation}
    for a suitable (dimensionless) physical quantity $d$ which is therefore independent of $u$.  
    
    \item{The choice of the physical dimensionless quantity $d$ is not unique. A candidate choice for it is the normalized proper distance from the center of the BH \cite{Bonanno:2000ep} 
    \begin{equation}
    d(z) := \frac{1}{\ell_\mathrm{P}} \int_0^{z \ell_\mathrm{P}} \sqrt{|\dd s^2|} =  \int_0^z \frac{\dd z^\prime}{\sqrt{|f(z^\prime)|}} \ .
    \end{equation}}
 The first integral is understood for fixed values of the angular coordinates ($\theta, \phi$). Notice that, as remarked above, since $d(z)$ cannot depend on $u$ this implies a constrain for   $\displaystyle{ f(z,\frac{u}{\ell_\mathrm{P}})}$ which restricts the $u$ dependence of $\displaystyle{g(d,\frac{u}{\ell_\mathrm{P}})}$. We note that quantum improvements of the metric based on unphysical quantities such as the radial coordinate $z$ lead to quantum geometries depending on the specific choice of coordinates. This issue was discussed in detail in~\cite{Held:2021vwd}. 
    
    \item  At large proper distances from the BH the function $g$  approaches asymptotically unity.  At the quantum level we have, therefore, for $f(z)$: 
    \begin{equation}\label{metricf}
        f(z)=  1-\frac{2\chi}{z}\sum_{n=0}^{\infty} \frac{\Omega_n}{d(z)^{2n}}\, .
    \end{equation}
    The specific values of the dimensionless coefficients $\Omega_n(\frac{u}{\ell_\mathrm{P}})$ , with $\Omega_0=1$, are dictated by a given theory of quantum gravity. The $u$ dependence of the $\Omega_n$ coefficients is constrained by requiring physical quantities to be independent on this arbitrary scale.  
    The expansion in \eqref{metricf} is built to incorporate the fact that at large distances the metric must asymptotically approach the classical one \eqref{metricc}. The choice of even inverse powers of $d(z)$ comes from our expectation that this quantum metric emerges from a(n effective) quantum gravity action with only even powers of the derivatives. We have also neglected subleading logarithmic terms. The approach can be readily extended to include a different counting scheme if required by more general theories of quantum gravity. A different definition of the physical distance $d(z)$ leads to modified coefficients $\Omega_n$.  
   
    \item By construction \eqref{metricf} is an involved equation for $f(z)$ which we attack in a self consistent iterative manner{: we shall} add one order in $n$ of the series at a time and include the backreaction stemming from the corrected  $d(z)$ from the previous order. In practice, this procedure mimics the potential expansion of an effective quantum gravity action in local derivative operators. However, the overall approach does not rely on this interpretation and can therefore be further extended to include non-analytic terms, which we plan to explore  in the future.
 \end{enumerate}
 In the following subsection we start by considering the leading quantum correction.

\subsection{Leading Order Quantum Metric}
\begin{figure}[!ht]
        \centering
        \includegraphics[width=0.8\textwidth]{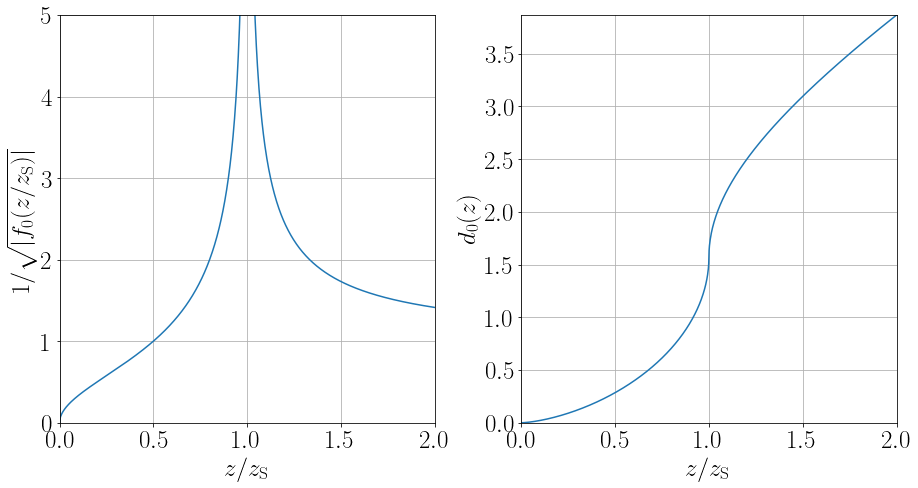}
\caption{Left panel: We plot the integrand of \eqref{eq: schwarz geodesic distance integrand},  {normalised to} the  dimensionless Schwarzschild radius of the BH $z_\mathrm{S}=2 \chi $. Right panel: the regular {proper} distance $d_0(z)$ given in ~\eqref{eq: schwarz geodesic distance}.}
\label{fig: schwarz geod distance}
\end{figure}

To determine the leading order quantum corrected function $f_1(z)$, we introduce the classical proper distance $d_0(z)$ given by
\begin{equation}
    d_0 (z)
     =\int_0^z \frac{\dd z^\prime}{\sqrt{\left|f_0(z^\prime)\right|}}
    =\int_0^z \frac{\dd z^\prime}{\sqrt{\left|1-\frac{2 \chi}{z^\prime}\right|}}\, .
    \label{eq: schwarz geodesic distance integrand}
\end{equation}
 Performing the integration we have 
\begin{equation}\label{eq: schwarz geodesic distance}
    d_0(z)=\begin{cases}\displaystyle
    \pi \chi - 2 \chi \tan^{-1} \sqrt{\frac{2 \chi}{z}-1}-\sqrt{z(2 \chi -z)}\, , & 0<z<2 \chi\, ,\\ \\ \displaystyle
    \pi \chi + 2\chi \tanh^{-1}\sqrt{1-\frac{2 \chi}{z}}+\sqrt{z(z-2 \chi)}\, ,  & 2 \chi < z < \infty\, .
    \end{cases}
\end{equation}
 The left panel of Fig.~\ref{fig: schwarz geod distance} is the graphical representation of the integrand in \eqref{eq: schwarz geodesic distance integrand} while the right panel represents  \eqref{eq: schwarz geodesic distance}. The integrand \eqref{eq: schwarz geodesic distance integrand} has an integrable singularity at $z_S = 2\chi$ yielding a regular {proper} distance.  
\\Near the center of the BH the distance function behaves like
\begin{equation}\label{eq: Bonanno small r expansion}
    d_0(z) \simeq \frac{2}{3}\frac{z^{3/2}}{\sqrt{2 \chi}}+ \mathcal{O}\left(\frac{z^{5/2}}{(2 \chi)^{3/2}}\right)\, ,
\end{equation}
 while a linear dependence is recovered at distances far from the horizon as shown in Fig.~\ref{fig: schwarz geod distance}. The leading order quantum corrected $f$ function reads 
\begin{equation}\label{eq: schw Z quantum corrected}
    f_1(z)=1- 2\frac{\chi}{z} \left[1 + \frac{\Omega_1}{d_0^2(z)}\right]\, .
\end{equation}

\begin{figure}[!ht]
    \centering
    \includegraphics[width=0.8\textwidth]{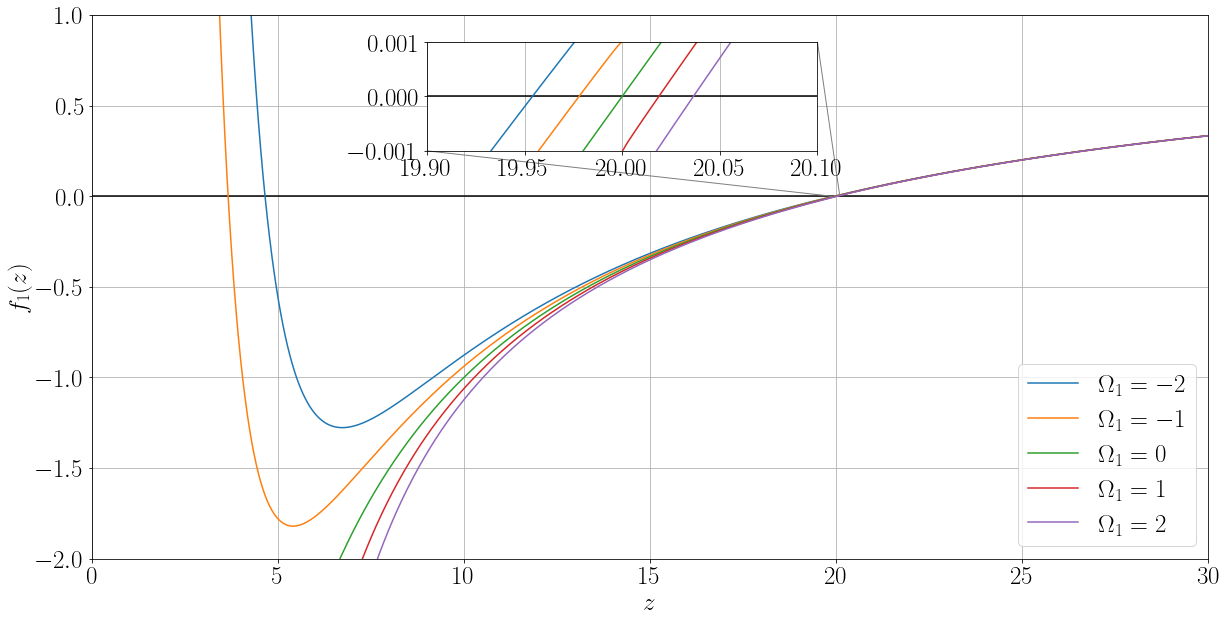}
    \caption{Plot of the quantum corrected function $f_1(z)$ given in~\eqref{eq: schw Z quantum corrected} as a function of the distance in Planck units $z$ and for different values of  $\Omega_1$ for a fixed mass ratio $\chi=10$.}
    \label{fig: schw metric f varying Omega}
\end{figure}
 
 \subsection{Quantum Horizons}\label{sec: horizons}
We are now ready to discuss the quantum corrections to the BH classical horizon starting with the zeros of $f_1 (z)$. Here, the location of the zeroes depends on the sign of the parameter $\Omega_1$ that leads to qualitatively different BH solutions. To remain general, we discuss both cases separately. We plot in Fig.~\ref{fig: schw metric f varying Omega} the function $f_1(z)$ for different values of $\Omega_1$ (while we fixed $\chi=10$). This plot shows two qualitatively very important results
\begin{itemize}
    \item[1.]{The position of the (external) horizon is a function of $\Omega_1$: positive values of $\Omega_1$ shift the zero of $f_1$ to larger values of $z$, while negative values of $\Omega_1$ move it to smaller values. However, in both cases, the effect is small compared to the classical position of the horizon $2\chi$, even for values of $|\Omega_1|\geq 1$.}
    \item [2.] {For negative values of $\Omega_1$, the function $f_1$ allows for a second zero in the physical region $z>0$, which can be interpreted as the formation of a new \emph{internal horizon}. The position of the latter depends much stronger on the numerical value of $\Omega_1$ than the corrections to the external horizon.}
\end{itemize}
In the following we shall discuss both effects more quantitatively, by treating separately the different signs of $\Omega_1$.

\subsubsection{\texorpdfstring{$\Omega_1>0$}{Omega positive}}
As shown in Fig.~\ref{fig: schw metric f varying Omega}, for positive values of $\Omega_1$ the function $f_1$ has a single zero for $z>0$ corresponding to a single horizon which can be expanded around the classical solution as follows: 
\begin{equation}
z_+ = 2\chi \left[1 +  \left(\frac{\Omega_1}{\pi^2\chi^2}\right)  + {\cal{O}} {\left(\frac{\Omega_1}{\pi^2\chi^2}\right)^{3/2}}\right]\equiv 2 \chi  \left[1 +  \alpha + \, {\mathcal{O}}\left(\alpha^{3/2} \right)\right] \, ,\qq{with}\alpha:=\frac{|\Omega_1|}{\pi^2\chi^2}\,,\label{eq: external horzion positive omega} 
\end{equation}
  We can render $\alpha$ arbitrarily small by increasing $\chi$ (the BH mass) for fixed $\abs{\Omega_1}$. While in practice the factor of $\pi^2$ in the definition of $\alpha$ in (\ref{eq: external horzion positive omega}) further suppresses the quantum corrections, we remark that $\pi\chi = d_0(2\chi)$ is in fact the classical distance of the classical BH horizon (see eq.~(\ref{eq: schwarz geodesic distance})). Therefore, the expansion in (\ref{eq: external horzion positive omega}) is organised in terms of physical quantities of the classical BH geometry. Naturally, we recover the classical horizon when we either switch off the quantum corrections or increase the BH mass (such that $\alpha\to 0$).  We discuss  the numerical range of validity of (\ref{eq: external horzion positive omega}) as a function of the BH mass in Appendix~\ref{App:RadiusConvergence} and higher order corrections in $\alpha$ in Appendix~\ref{App:NumericPosition}. As we shall demonstrate in the next subsection, the corrections to $z_+$ stemming from self consistently replacing $d_0(z)$ in $f_1(z)$ (see  \eqref{eq: schw Z quantum corrected}) with 
 \begin{equation}\label{d1 v1}
d_1(z)=\int_0^z \frac{\dd z^\prime}{\sqrt{\left|f_1(z^\prime)\right|}}\ ,
 \end{equation} 
  appear at ${\mathcal{O}}(\alpha^{3/2})$. Therefore, to this order in $\alpha$, all the quantum corrections are taken into account for the external horizon. The horizon location could depend on the unphysical scale $u$ through $\Omega_1$ which, however, to the current quantum order is constrained to be $u$ independent by requiring $d_1$ to be a physical quantity to the same order. This can   be seen from  \eqref{d1sol}. 
  
Overall, the horizon increases due to quantum corrections and these are further suppressed at large masses.

\subsubsection{\texorpdfstring{$\Omega_1<0$}{Omega negative}}
In this case the quantum corrected  external horizon reads: 
\begin{align}
z_+ =  2 \chi  \left[1 -  \alpha + \, {\mathcal{O}}\left(\alpha^{3/2} \right)  \right]\ ,\label{eq: external horizon negative omega}
\end{align}
where $\alpha$ is defined as in (\ref{d1 v1}). The external horizon, now, decreases due to quantum corrections, while the other remarks made for the $\Omega_1$ positive case still apply.
For negative values of $\Omega_1$, as shown in Fig.~\ref{fig: schw metric f varying Omega}, an internal horizon forms  at the position
\begin{align}
z_-=\chi\left(\frac{9\pi}{2}\right)^{1/3}\,\left[\alpha^{1/3}+\frac{1}{5}\left(\frac{\pi^2}{6}\right)^{1/3}\,\alpha^{2/3}+\frac{61}{700}\,\left(\frac{3\pi^4}{4}\right)^{1/3}\,\alpha+\mathcal{O}(\alpha^{4/3})\right] \ .
\end{align}
Clearly, the existence of the internal horizon has a quantum nature {and strongly depends} on the underlying theory of quantum gravity. 

\begin{figure}[!ht]
    \centering
    \includegraphics[width=0.85\textwidth]{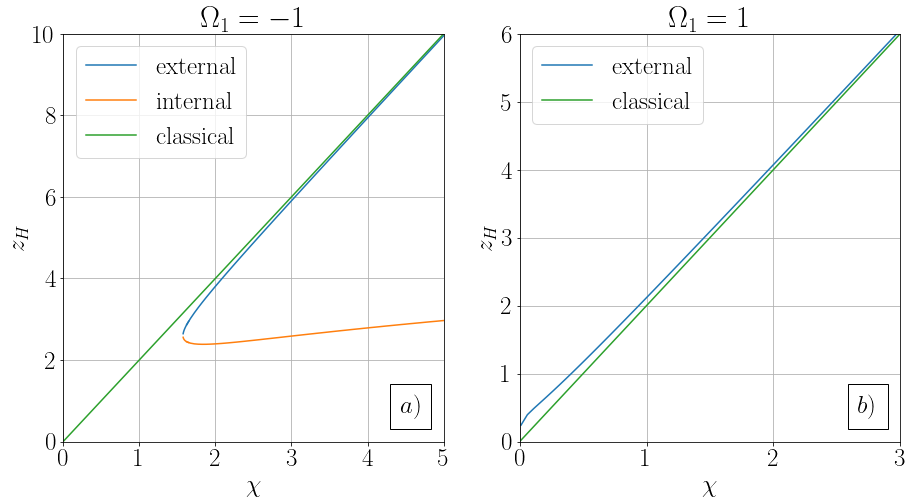}
    \caption{Black hole event horizons  indicated by $z_H$ corresponding to either the classical (straight line), $z_+$ and $z_-$, as functions of the mass ratio $\chi$ for the value of $\Omega_1=-1$ ($a$) and $\Omega_1=1$ ($b$). The values are found solving numerically $f_1(z)=0$.}
    \label{fig: sch correctwarz horizons}
\end{figure}
We display the zeroes of $f_1(z)$ in  Fig.~\ref{fig: sch correctwarz horizons} as function of $\chi$ and observe that the internal horizon is less dependent on this parameter when compared to the external one. Furthermore, for masses close to the Planck value the two horizons merge leading to an extremality condition that is analytically approximated to be:  
\begin{equation}
    \chi^{\rm ext} \simeq \frac{16\sqrt{-\Omega_1}}{\pi^2}\, .
    \label{S-extremal}
\end{equation} 
The ultimate fate for the existence of the internal horizon depends, as we shall see, on the size and sign of the higher order corrections.

\subsubsection{Conformal Diagrams}
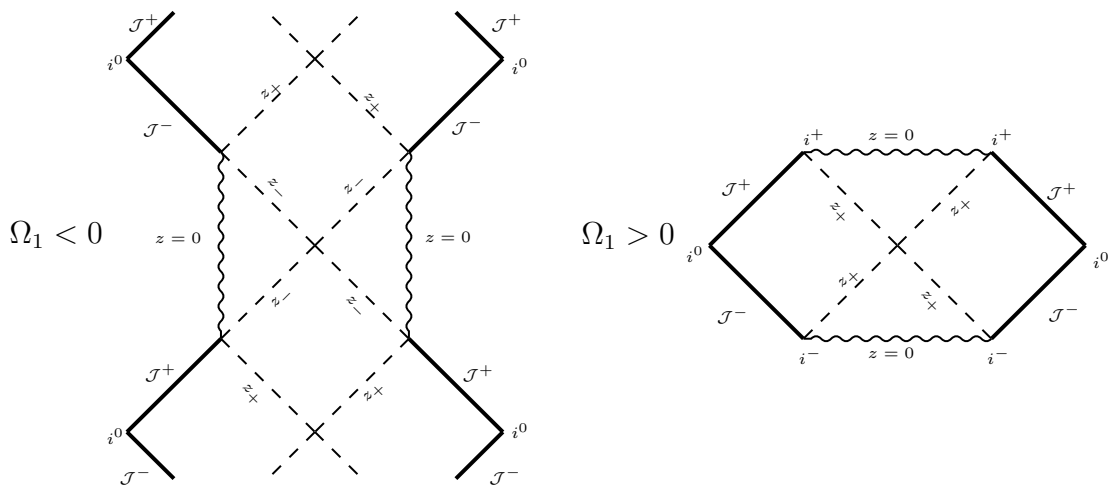
\begin{figure}[htbp]
\centering
\tikzset{every picture/.style={line width=0.75pt}} 

\begin{tikzpicture}[x=0.75pt,y=0.75pt,yscale=-1,xscale=1]

\draw [line width=1.5]    (146.53,100.95) -- (99.61,54.04) ;
\draw  [dash pattern={on 4.5pt off 4.5pt}]  (146.53,100.95) -- (193.44,54.04) ;
\draw  [dash pattern={on 4.5pt off 4.5pt}]  (193.44,54.04) -- (240.35,100.95) ;
\draw [line width=1.5]    (240.35,100.95) -- (287.26,54.04) ;
\draw [line width=1.5]    (99.61,54.04) -- (123.07,30.58) ;
\draw  [dash pattern={on 4.5pt off 4.5pt}]  (193.44,54.04) -- (169.98,30.58) ;
\draw  [dash pattern={on 4.5pt off 4.5pt}]  (193.44,54.04) -- (216.9,30.58) ;
\draw [line width=1.5]    (287.26,54.04) -- (263.81,30.58) ;
\draw  [dash pattern={on 4.5pt off 4.5pt}]  (146.53,100.95) -- (193.44,147.86) ;
\draw  [dash pattern={on 4.5pt off 4.5pt}]  (193.44,147.86) -- (240.35,100.95) ;
\draw  [dash pattern={on 4.5pt off 4.5pt}]  (193.44,147.86) -- (146.53,194.78) ;
\draw  [dash pattern={on 4.5pt off 4.5pt}]  (193.44,147.86) -- (240.35,194.78) ;
\draw    (146.53,100.95) .. controls (148.2,102.62) and (148.2,104.28) .. (146.53,105.95) .. controls (144.86,107.62) and (144.86,109.28) .. (146.53,110.95) .. controls (148.2,112.62) and (148.2,114.28) .. (146.53,115.95) .. controls (144.86,117.62) and (144.86,119.28) .. (146.53,120.95) .. controls (148.2,122.62) and (148.2,124.28) .. (146.53,125.95) .. controls (144.86,127.62) and (144.86,129.28) .. (146.53,130.95) .. controls (148.2,132.62) and (148.2,134.28) .. (146.53,135.95) .. controls (144.86,137.62) and (144.86,139.28) .. (146.53,140.95) .. controls (148.2,142.62) and (148.2,144.28) .. (146.53,145.95) .. controls (144.86,147.62) and (144.86,149.28) .. (146.53,150.95) .. controls (148.2,152.62) and (148.2,154.28) .. (146.53,155.95) .. controls (144.86,157.62) and (144.86,159.28) .. (146.53,160.95) .. controls (148.2,162.62) and (148.2,164.28) .. (146.53,165.95) .. controls (144.86,167.62) and (144.86,169.28) .. (146.53,170.95) .. controls (148.2,172.62) and (148.2,174.28) .. (146.53,175.95) .. controls (144.86,177.62) and (144.86,179.28) .. (146.53,180.95) .. controls (148.2,182.62) and (148.2,184.28) .. (146.53,185.95) .. controls (144.86,187.62) and (144.86,189.28) .. (146.53,190.95) -- (146.53,194.78) -- (146.53,194.78) ;
\draw    (240.35,100.95) .. controls (242.02,102.62) and (242.02,104.28) .. (240.35,105.95) .. controls (238.68,107.62) and (238.68,109.28) .. (240.35,110.95) .. controls (242.02,112.62) and (242.02,114.28) .. (240.35,115.95) .. controls (238.68,117.62) and (238.68,119.28) .. (240.35,120.95) .. controls (242.02,122.62) and (242.02,124.28) .. (240.35,125.95) .. controls (238.68,127.62) and (238.68,129.28) .. (240.35,130.95) .. controls (242.02,132.62) and (242.02,134.28) .. (240.35,135.95) .. controls (238.68,137.62) and (238.68,139.28) .. (240.35,140.95) .. controls (242.02,142.62) and (242.02,144.28) .. (240.35,145.95) .. controls (238.68,147.62) and (238.68,149.28) .. (240.35,150.95) .. controls (242.02,152.62) and (242.02,154.28) .. (240.35,155.95) .. controls (238.68,157.62) and (238.68,159.28) .. (240.35,160.95) .. controls (242.02,162.62) and (242.02,164.28) .. (240.35,165.95) .. controls (238.68,167.62) and (238.68,169.28) .. (240.35,170.95) .. controls (242.02,172.62) and (242.02,174.28) .. (240.35,175.95) .. controls (238.68,177.62) and (238.68,179.28) .. (240.35,180.95) .. controls (242.02,182.62) and (242.02,184.28) .. (240.35,185.95) .. controls (238.68,187.62) and (238.68,189.28) .. (240.35,190.95) -- (240.35,194.78) -- (240.35,194.78) ;
\draw  [dash pattern={on 4.5pt off 4.5pt}]  (146.53,194.78) -- (193.44,241.69) ;
\draw  [dash pattern={on 4.5pt off 4.5pt}]  (240.35,194.78) -- (193.44,241.69) ;
\draw [line width=1.5]    (146.53,194.78) -- (99.61,241.69) ;
\draw [line width=1.5]    (240.35,194.78) -- (287.26,241.69) ;
\draw [line width=1.5]    (99.61,241.69) -- (123.07,265.15) ;
\draw  [dash pattern={on 4.5pt off 4.5pt}]  (193.44,241.69) -- (169.98,265.15) ;
\draw  [dash pattern={on 4.5pt off 4.5pt}]  (193.44,241.69) -- (216.9,265.15) ;
\draw [line width=1.5]    (287.26,241.69) -- (263.81,265.15) ;
\draw [line width=1.5]    (531.11,100.95) -- (578.03,147.86) ;
\draw [line width=1.5]    (578.03,147.86) -- (531.11,194.78) ;
\draw  [dash pattern={on 4.5pt off 4.5pt}]  (484.2,147.86) -- (531.11,194.78) ;
\draw  [dash pattern={on 4.5pt off 4.5pt}]  (531.11,100.95) -- (484.2,147.86) ;
\draw  [dash pattern={on 4.5pt off 4.5pt}]  (437.29,100.95) -- (484.2,147.86) ;
\draw  [dash pattern={on 4.5pt off 4.5pt}]  (484.2,147.86) -- (437.29,194.78) ;
\draw [line width=1.5]    (390.37,147.86) -- (437.29,194.78) ;
\draw [line width=1.5]    (437.29,100.95) -- (390.37,147.86) ;
\draw    (531.11,100.95) .. controls (529.44,102.62) and (527.78,102.62) .. (526.11,100.95) .. controls (524.44,99.28) and (522.78,99.28) .. (521.11,100.95) .. controls (519.44,102.62) and (517.78,102.62) .. (516.11,100.95) .. controls (514.44,99.28) and (512.78,99.28) .. (511.11,100.95) .. controls (509.44,102.62) and (507.78,102.62) .. (506.11,100.95) .. controls (504.44,99.28) and (502.78,99.28) .. (501.11,100.95) .. controls (499.44,102.62) and (497.78,102.62) .. (496.11,100.95) .. controls (494.44,99.28) and (492.78,99.28) .. (491.11,100.95) .. controls (489.44,102.62) and (487.78,102.62) .. (486.11,100.95) .. controls (484.44,99.28) and (482.78,99.28) .. (481.11,100.95) .. controls (479.44,102.62) and (477.78,102.62) .. (476.11,100.95) .. controls (474.44,99.28) and (472.78,99.28) .. (471.11,100.95) .. controls (469.44,102.62) and (467.78,102.62) .. (466.11,100.95) .. controls (464.44,99.28) and (462.78,99.28) .. (461.11,100.95) .. controls (459.44,102.62) and (457.78,102.62) .. (456.11,100.95) .. controls (454.44,99.28) and (452.78,99.28) .. (451.11,100.95) .. controls (449.44,102.62) and (447.78,102.62) .. (446.11,100.95) .. controls (444.44,99.28) and (442.78,99.28) .. (441.11,100.95) -- (437.29,100.95) -- (437.29,100.95) ;
\draw    (531.11,194.78) .. controls (529.44,196.45) and (527.78,196.45) .. (526.11,194.78) .. controls (524.44,193.11) and (522.78,193.11) .. (521.11,194.78) .. controls (519.44,196.45) and (517.78,196.45) .. (516.11,194.78) .. controls (514.44,193.11) and (512.78,193.11) .. (511.11,194.78) .. controls (509.44,196.45) and (507.78,196.45) .. (506.11,194.78) .. controls (504.44,193.11) and (502.78,193.11) .. (501.11,194.78) .. controls (499.44,196.45) and (497.78,196.45) .. (496.11,194.78) .. controls (494.44,193.11) and (492.78,193.11) .. (491.11,194.78) .. controls (489.44,196.45) and (487.78,196.45) .. (486.11,194.78) .. controls (484.44,193.11) and (482.78,193.11) .. (481.11,194.78) .. controls (479.44,196.45) and (477.78,196.45) .. (476.11,194.78) .. controls (474.44,193.11) and (472.78,193.11) .. (471.11,194.78) .. controls (469.44,196.45) and (467.78,196.45) .. (466.11,194.78) .. controls (464.44,193.11) and (462.78,193.11) .. (461.11,194.78) .. controls (459.44,196.45) and (457.78,196.45) .. (456.11,194.78) .. controls (454.44,193.11) and (452.78,193.11) .. (451.11,194.78) .. controls (449.44,196.45) and (447.78,196.45) .. (446.11,194.78) .. controls (444.44,193.11) and (442.78,193.11) .. (441.11,194.78) -- (437.29,194.78) -- (437.29,194.78) ;

\draw (246.77,139.4) node [anchor=north west][inner sep=0.75pt]  [font=\tiny]  {$z=0$};
\draw (214.67,224.01) node [anchor=north west][inner sep=0.75pt]  [font=\tiny,rotate=-316.26]  {$z_{+}$};
\draw (159.47,214.55) node [anchor=north west][inner sep=0.75pt]  [font=\tiny,rotate=-45.28]  {$z_{+}$};
\draw (211.1,170.34) node [anchor=north west][inner sep=0.75pt]  [font=\tiny,rotate=-45.28]  {$z_{-}$};
\draw (168.76,176.89) node [anchor=north west][inner sep=0.75pt]  [font=\tiny,rotate=-314.64]  {$z_{-}$};
\draw (205,120) node [anchor=north west][inner sep=0.75pt]  [font=\tiny,rotate=-314.64]  {$z_{-}$};
\draw (171.9,110.89) node [anchor=north west][inner sep=0.75pt]  [font=\tiny,rotate=-45.28]  {$z_{-}$};
\draw (163.43,72.08) node [anchor=north west][inner sep=0.75pt]  [font=\tiny,rotate=-316.26]  {$z_{+}$};
\draw (220,68) node [anchor=north west][inner sep=0.75pt]  [font=\tiny,rotate=-45.28]  {$z_{+}$};
\draw (290,235.74) node [anchor=north west][inner sep=0.75pt]  [font=\tiny]  {$i^{0}$};
\draw (290,52.3) node [anchor=north west][inner sep=0.75pt]  [font=\tiny]  {$i^{0}$};
\draw (88 ,238.59) node [anchor=north west][inner sep=0.75pt]  [font=\tiny]  {$i^{0}$};
\draw (88 ,50.56) node [anchor=north west][inner sep=0.75pt]  [font=\tiny]  {$i^{0}$};
\draw (112.04,139.85) node [anchor=north west][inner sep=0.75pt]  [font=\tiny]  {$z=0$};
\draw (39.53,134.25) node [anchor=north west][inner sep=0.75pt]    {$\Omega_1 < 0$};
\draw (265.57,206.38) node [anchor=north west][inner sep=0.75pt]  [font=\tiny]  {$\mathcal{J}^{+}$};
\draw (279.89,29.99) node [anchor=north west][inner sep=0.75pt]  [font=\tiny]  {$\mathcal{J}^{+}$};
\draw (107.03,206) node [anchor=north west][inner sep=0.75pt]  [font=\tiny]  {$\mathcal{J}^{+}$};
\draw (98.72,29.96) node [anchor=north west][inner sep=0.75pt]  [font=\tiny]  {$\mathcal{J}^{+}$};
\draw (260,82.02) node [anchor=north west][inner sep=0.75pt]  [font=\tiny]  {$\mathcal{J}^{-}$};
\draw (278.92,257.94) node [anchor=north west][inner sep=0.75pt]  [font=\tiny]  {$\mathcal{J}^{-}$};
\draw (105.95,83.11) node [anchor=north west][inner sep=0.75pt]  [font=\tiny]  {$\mathcal{J}^{-}$};
\draw (94.36,258.12) node [anchor=north west][inner sep=0.75pt]  [font=\tiny]  {$\mathcal{J}^{-}$};
\draw (325,134.6) node [anchor=north west][inner sep=0.75pt]    {$\Omega_1  >0$};
\draw (557.95,175.84) node [anchor=north west][inner sep=0.75pt]  [font=\tiny]  {$\mathcal{J}^{-}$};
\draw (556.87,114.86) node [anchor=north west][inner sep=0.75pt]  [font=\tiny]  {$\mathcal{J}^{+}$};
\draw (580.24,150.04) node [anchor=north west][inner sep=0.75pt]  [font=\tiny]  {$i^{0}$};
\draw (507.76,130.77) node [anchor=north west][inner sep=0.75pt]  [font=\tiny,rotate=-316.26]  {$z_{+}$};
\draw (498.13,167.79) node [anchor=north west][inner sep=0.75pt]  [font=\tiny,rotate=-45.28]  {$z_{+}$};
\draw (452.66,122.8) node [anchor=north west][inner sep=0.75pt]  [font=\tiny,rotate=-45.28]  {$z_{+}$};
\draw (451.85,166.68) node [anchor=north west][inner sep=0.75pt]  [font=\tiny,rotate=-316.26]  {$z_{+}$};
\draw (394.67,112.81) node [anchor=north west][inner sep=0.75pt]  [font=\tiny]  {$\mathcal{J}^{+}$};
\draw (377.31,146.96) node [anchor=north west][inner sep=0.75pt]  [font=\tiny]  {$i^{0}$};
\draw (392.5,177.3) node [anchor=north west][inner sep=0.75pt]  [font=\tiny]  {$\mathcal{J}^{-}$};
\draw (468,88) node [anchor=north west][inner sep=0.75pt]  [font=\tiny]  {$z=0$};
\draw (468,198) node [anchor=north west][inner sep=0.75pt]  [font=\tiny]  {$z=0$};
\draw (529.57,88) node [anchor=north west][inner sep=0.75pt]  [font=\tiny]  {$i^{+}$};
\draw (435.57,88) node [anchor=north west][inner sep=0.75pt]  [font=\tiny]  {$i^{+}$};
\draw (527.57,198) node [anchor=north west][inner sep=0.75pt]  [font=\tiny]  {$i^{-}$};
\draw (433.57,198) node [anchor=north west][inner sep=0.75pt]  [font=\tiny]  {$i^{-}$};

\end{tikzpicture}

\caption{Conformal diagrams of the quantum corrected space-time described by a metric given by the function in eq.~\eqref{eq: schw Z quantum corrected} for $\Omega_1<0$ (left) and $\Omega_1>0$ (right). The case of negative $\Omega_1$ corresponds to a BH with two distinct event horizons. The notation is the following: $\mathcal{J^+}$ ($\mathcal{J^-}$) is the future (past) null infinity, $i^+$ ($i^-$) is the future (past) timelike infinity and $i^0$ is the spatial infinity, while $z=0,\ z_+,\ z_-$ are respectively the singularity at the origin, the external  event horizon and the internal one.}
\label{fig: conformal diagrams}
\end{figure}

The global properties of the quantum space-time described by the metric with the $f$ function in eq.~\eqref{eq: schw Z quantum corrected} can be neatly summarised via Penrose's diagrams \cite{Carroll:2004st,Townsend:1997ku} shown in Fig.~\ref{fig: conformal diagrams}. 
The positive  $\Omega_1$ case can be summarised as similar to the classical Schwarzschild one with a spacelike singularity while the BH horizon is just slightly larger. The conformal diagram relative to the maximal extension of this space-time is shown in the right panel of Fig.~\ref{fig: conformal diagrams}, which is the Szekeres-Kruskal conformal diagram \cite{Carroll:2004st,Townsend:1997ku}.\\
Since for negative $\Omega_1$ the quantum BH has two event horizons, its space-time structure qualitatively resembles the classical Reissner-Nordstr\"om (RN) one \cite{Carroll:2004st,Townsend:1997ku,Bonanno:2000ep}. To better appreciate the differences we note that, at large distances, the RN dependence on the charge decreases as $z^{-2}$ while for the quantum corrected one it goes as $z^{-3}$ in terms of the quantum effects. The qualitative conformal diagram of the maximal extension of such a space-time is  therefore still expected to be of the form given in the left panel of Fig.~\ref{fig: conformal diagrams}.

As it is shown, the singularity at the origin is a timelike one and the two horizons $z_+$ and $z_-$ can be crossed by a timelike infalling observer who can reach multiple space-times. The case of an extremal BH is not shown and it can be found in the literature as its conformal diagram, as stated before, is analogous to the classical RN one.\\

\subsection{Quantum Proper Distance}  \label{QGD}
Even including only the leading quantum corrections by restricting to an effective second-derivative action (and thus truncating the series (\ref{metricf}) at $n=1$), eq.~(\ref{eq: schw Z quantum corrected}) is only an approximation, since it contains the classical proper distance $d_0$. Self consistency requires to {include} the impact of the quantum corrected geodesic distance, previously also indicated as proper distance, on the quantum $f$ function given in \eqref{eq: schw Z quantum corrected} by substituting $d_0$  
with 
\begin{equation}\displaystyle
d_1(z) = \int_0^z \frac{\dd z^\prime}{\sqrt{\left|1- \frac{2\chi}{z^\prime} \left[1 + \frac{\Omega_1}{d_0^2(z^\prime)}\right]\right|}}\, .
    \label{eq: quantum geodesic distance integrand}
\end{equation} 

\begin{figure}[!t]
    \centering
    \includegraphics[width=.9\textwidth]{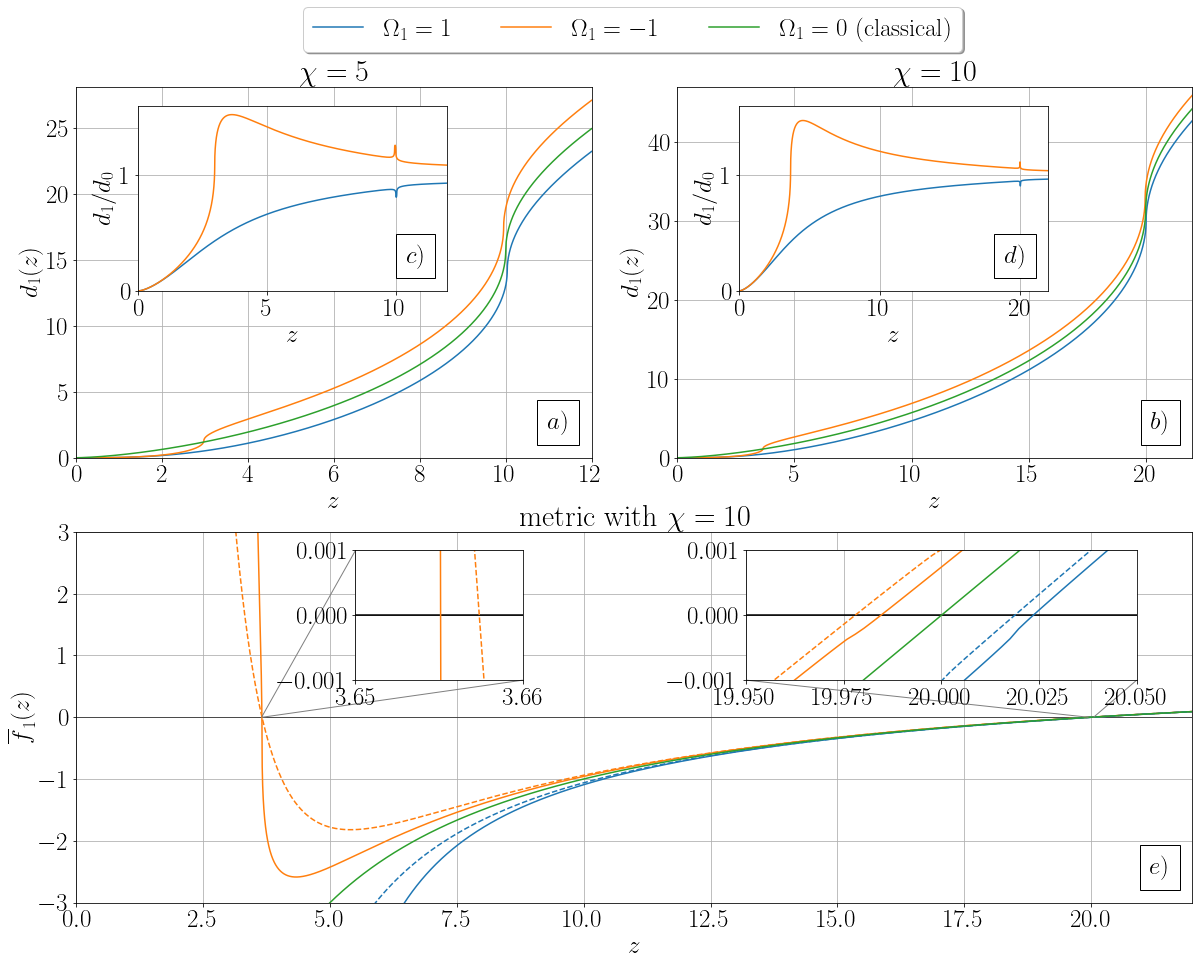}
    \caption{In the panels $a)$ and $b)$ we plot the dimensionless quantum corrected proper distance $d_1$ for $\Omega_1=1$ and $\Omega_1=-1$, and for two different values of the mass ratio $\chi=$ 5 ($a$), 10 ($b$). We also plot $d_0$ corresponding to the classical proper distance ($\Omega_1=0$). The ratio $d_1/d_0$ is displayed in  the inserts $c)$ and $d)$.   In the bottom panel $e)$ the solid lines (blue for $\Omega_1>0$ and orange for  $\Omega_1<0$) correspond to the improved quantum function $\bar{f}_1(z)$ while the dotted lines correspond to the unimproved $f_1(z)$. The green solid line represents the classical function $f_0(z)$.}
    \label{fig: fofrgeodesic}
\end{figure}
We  therefore obtain the quantum self improved $f$ function\footnote{The procedure is straightforwardly generalized when considering higher order quantum corrections, as we shall see below.}
\begin{equation} \label{eq: schw Z quantum geodesic improved}
     \bar{f}_1(z)= 1- 2\frac{\chi}{z} \left[1 + \frac{\Omega_1}{d_1^2(z)}\right]\, .
\end{equation}

\begin{figure}[!ht]
    \centering
    \includegraphics[width=0.85\textwidth]{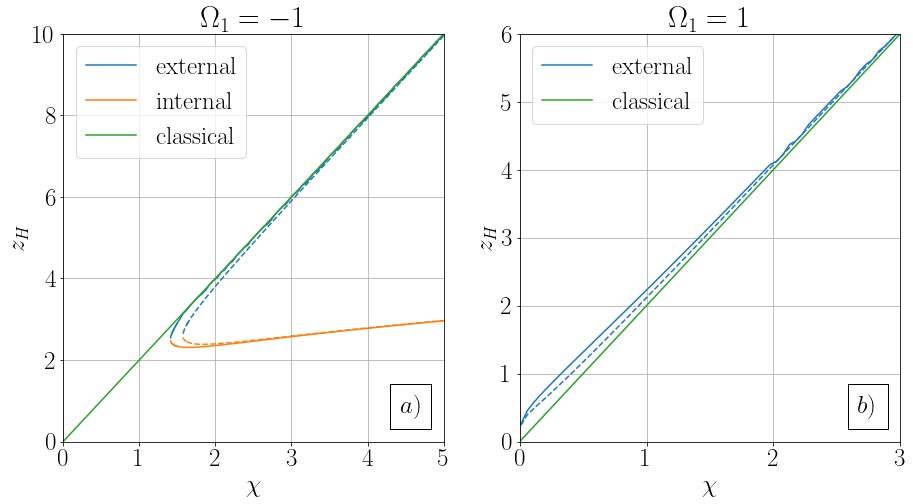}
    \caption{In the subplot $a)$ we set $\Omega_1=-1$ and we represent the external and internal event horizon dimensionless radii $z_H$ as a function of the mass ratio $\chi$ in the case of a space-time with back-reaction on $f$ described by the function in eq. \eqref{eq: schw Z quantum geodesic improved} (solid lines) and compare them with the ones that come from eq. \eqref{eq: schw Z quantum corrected} (dashed lines). In subplot $b)$ we study the case in which $\Omega_1=1$ and plot the event horizon radius as $\chi$ varies and compare it with the one without back-reaction on the $f$ discussion. Eventual oscillations in the plots are due to numerical errors.}
    \label{fig: horizons quantum geodesic}
\end{figure}

For $\Omega_1<0$, one observes that $d_1$ remains smaller than the classical {$d_0$} for values of $z$ smaller than the internal horizon while it is larger for any other value of $z$. Nevertheless the qualitative behaviour of the quantum distance mimics the classical geodesic one. A similar analysis for $\Omega_1>0$ is simplified by the fact that only the external horizon is present. Here  the quantum corrected {proper} distance $d_1$ again follows the behaviour of the classical distance. 
In panels $a)$ and $b)$ of Fig.~\ref{fig: fofrgeodesic} we display the classical and quantum {proper distances}. The two plots correspond to two different values of the BH mass.  We also observe that near the origin the quantum corrected geodesic distance approaches zero faster than the classical one, specifically it goes as $z^{3}$ rather than $z^{3/2}$. For completeness we plot the ratio of $d_1/d_0$ as function of $z$ in the panels 
$c)$ and $d)$ of Fig.~\ref{fig: fofrgeodesic} corresponding again to two different choices of the BH mass.

To acquire a  general understanding of the effects of the improved results for the function $\bar{f}_1$ of \eqref{eq: schw Z quantum geodesic improved} we plot it in Fig.~\ref{fig: fofrgeodesic}.  The solid green line corresponds to the classical function $f_0$, the orange and blue to the different signs of $\Omega_1$ taking into account the quantum corrected proper distance. The dashed curves correspond to the unimproved quantum  $f$ function obtained via the classical proper distance. As we had anticipated earlier, the location of the horizons, shown in Fig.~\ref{fig: horizons quantum geodesic}, are marginally affected by the improvement due to the quantum proper distance back-reaction. Specifically, the quantum proper distance improvement appear, for the external horizon to the  ${\cal O}(\alpha^{3/2})$, and for the internal one (for $\Omega_1$ negative) a numerical investigation suggests that the improvement appears beyond the order  ${\cal O}(\alpha)$.\\

\section{Higher Order Quantum Corrections} \label{HOC}
 {After having treated the leading quantum corrections in the previous section, we shall} now discuss the procedure to self consistently consider higher order quantum corrections.  {That is, we consider higher corrections to the $f$ function, but still truncate the sum in eq.~(\ref{metricf}) at a finite $n$.\footnote{{Notice that by keeping $n$ finite allows us to avoid questions about the radius of convergence of the sum in eq.~(\ref{metricf}). The latter is equivalent to the question whether the underlying theory of quantum gravity allows for a non-perturbative definition beyond a (perturbative) effective approach.}} In this way,} we assume that the underlying quantum corrected gravitational theory can be approximated via a local effective action featuring higher derivative operators {up to order $2n$}. {Thus, we} expect the resulting $f$ function to assume the form 
\begin{equation}
\bar{f}_n(z)=1-\frac{2\chi}{z}\left(1+\frac{\Omega_1}{\bar{d}_{n}^2(z)} +\frac{\Omega_2}{\bar{d}_{n}^4(z)} + \cdots +\frac{\Omega_n}{\bar{d}_{n}^{2n}(z)}\right) \ , \label{eq: n order barf} 
\end{equation}
with 
\begin{equation}\displaystyle
\bar{d}_n(z) = \int_0^z \frac{\dd z^\prime}{\sqrt{\left| \bar{f}_n(z^\prime)\right|}}\, .
    \label{eq: full quantum geodesic distance integrand n}
\end{equation} 
To be able to compute this quantity in an iterative manner we approximate it via
\begin{equation}\displaystyle
{d}_n(z) = \int_0^z \frac{\dd z^\prime}{\sqrt{\left| {f}_n(z^\prime)\right|}} \equiv\int_0^z \frac{\dd z^\prime}{\sqrt{\left|1-\frac{2\chi}{z^\prime}\left(1+\frac{\Omega_1}{d_{n-1}^2(z\prime)} +\frac{\Omega_2}{d_{n-1}^4(z\prime)} + \cdots +\frac{\Omega_n}{d_{n-1}^{2n}(z\prime)}\right)  \right|}}\, .
    \label{eq: quantum geodesic distance integrand n}
\end{equation} 
where $d_{n-1}$ is the quantum corrected {proper} distance at order $n-1$.  

{Even for $n>1$, }there are two limits where the full behavior of $d_n$ as function of $z$  {can be studied}:
\begin{itemize}
\item[1.] {{\bf asymptotically large distance:}}   {Far away from the black hole} $d_n$ approaches $z$.  {This limiting behaviour is crucial for the self-consistency of our approach: indeed, it is required} to match the  effective coefficients $\Omega_n$ to specific predictions from a given  underlying quantum gravity theory. {The universality of this limit
\begin{align}
    &\lim_{z\to \infty} d_n(z)=\lim_{z\to \infty}d_0(z)\,,&&\forall n\geq 0\,,
\end{align}
ensures that the coefficients $\Omega_{i\leq n}$ can be defined in a consistent fashion, independent of the order $n$.}
\item[2.]  {{\bf distances close to the center of the BH:}} Here the the dominant term in the integrand of \eqref{eq: quantum geodesic distance integrand n} is the last term in the denominator. This allows us to deduce, up to a multiplicative number, the following  relation 
\begin{equation} \label{recursion}
    \lim_{z \to 0} \frac{d_n}{d_0} \sim \lim_{z \to 0}      \frac{d^n_{n-1} }{\sqrt{\abs{\Omega_n}}}\ ,
\end{equation}
  with $d_0$ computed near the origin of the BH and therefore given by the first term in  \eqref{eq: Bonanno small r expansion}. { {Iteratively, this relation suggests
\begin{align}
\lim_{z\to 0} d_n\sim\lim_{z\to 0}\frac{(d_0)^{e\,\Gamma(n+1,1)}}{\sqrt{\prod_{i=1}^n|\Omega_i|^{n!/i!}}} 
\, ,
\end{align}
where $\Gamma(n+1,1)$ is the incomplete Gamma function (and we have implicitly assumed that all $\Omega_{i=1,\ldots,n}\neq 0$). 
}} This implies that near the origin, at each given order in $n$, the truncated physical quantum distance  approaches zero extremely fast. 
\end{itemize}
Now, we consider the explicit case of $n=2$  to learn how it 
affects our previous results \footnote{ {More details on the anaylsis can be found in Appendix~\ref{App:SecondOrderCorr}.}}
\begin{align}\label{eq: second order f}
f_2(z)&=1-\frac{2\chi}{z}\left(1+\frac{\Omega_1}{d_1^2(z)} +\frac{\Omega_2}{d_1^4(z)}\right) \ . 
\end{align}
Using \eqref{recursion} we have that 
\begin{equation}
  \lim_{z \to 0} {d_1} \sim \lim_{z \to 0} \frac{d_0^2}{\sqrt{\abs{\Omega_1}}} \ , \qquad {\rm and} \qquad  \lim_{z \to 0} {d_2} \sim \lim_{z \to 0} \frac{d_0^5}{{\abs{\Omega_2}^\frac{1}{2}}\abs{\Omega_1}} \ .\label{Scalingd1d2}
\end{equation}
Therefore, near the origin the corrections stemming from $d_2$  do not affect  $f_2$ since they enter at higher orders.\footnote{ {These limits are also recovered in a slightly different fashion in Appendix~\ref{App:SecondOrderCorr}.}} In Fig.~\ref{fig: fofrgeodesic2} we compare the quantum function $f_2$ with the improved $\bar{f}_1$, for different values of  $\Omega_1 = \pm 1$ and $\Omega_2 = -0.5,0.2,0.5$. Of course, when comparing with $\bar{f}_1$ we use the same values of $\Omega_1$ appearing in $f_2$.

\begin{figure}[!ht]
    \centering
    \includegraphics[width=.8\textwidth]{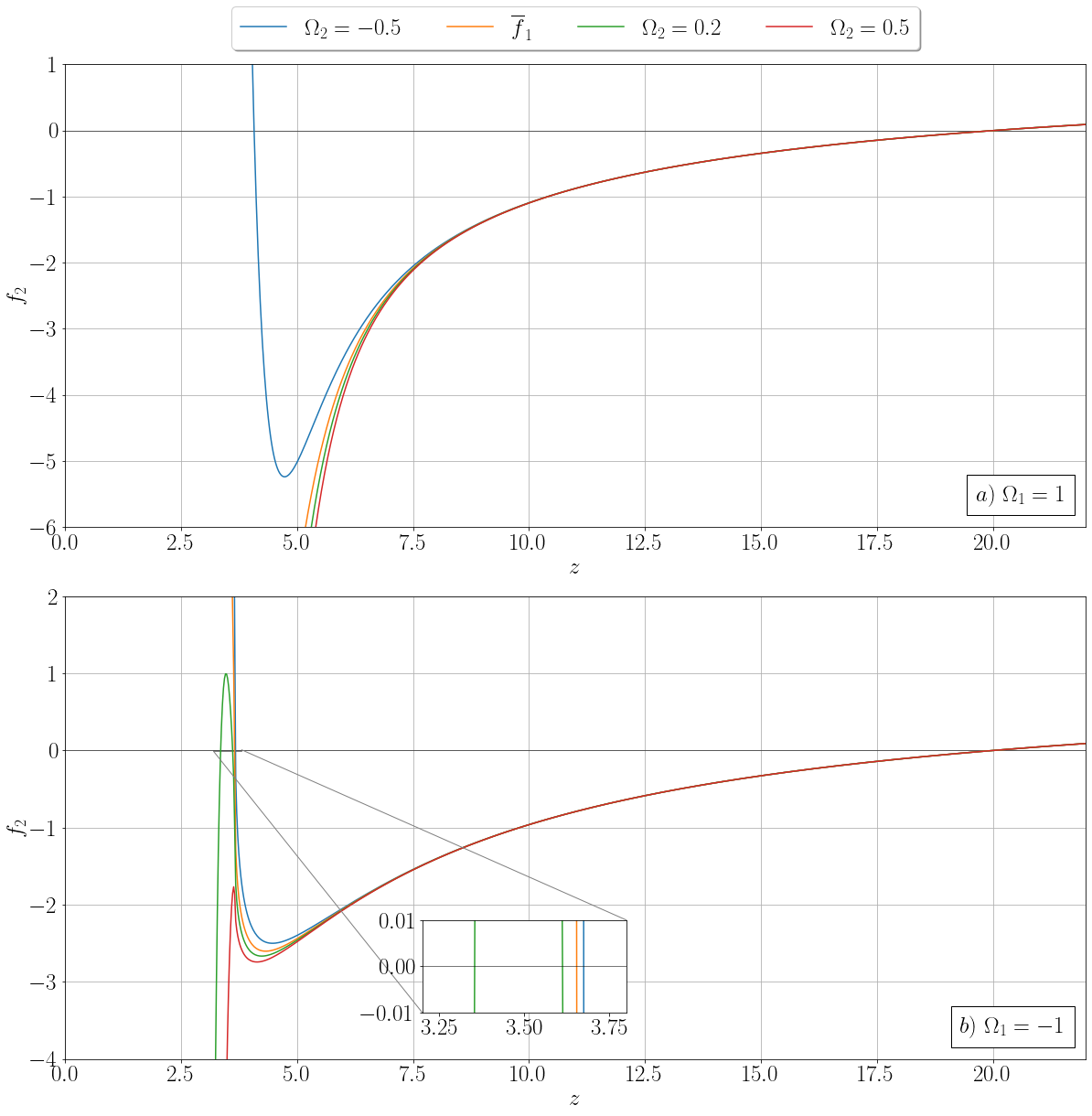}
    \caption{In top panel $a)$ we present $f_2$ for $\Omega_1=1$ and three values of $\Omega_2$ as shown in the legend on the top, with the improved function $\bar{f}_1$ displayed as a solid orange line. In the bottom panel $b)$ is as in $a)$ but with  $\Omega_1=-1$. In both plots we have set the BH mass relative to the Planck one to $\chi=10$. }
    \label{fig: fofrgeodesic2}
\end{figure}
We observe that, for both signs of $\Omega_1$, the external horizon expansion works extremely well yielding   {only minor} corrections stemming from $f_2$ when keeping $\Omega_2$ of order unity. This result is confirmed by the analytic expression of the external horizon in the expansion in the inverse BH mass which reads, for either positive or negative values of $\Omega_{1,2}$: 
\begin{equation}
    z_+ = 2\chi\left[ 1 + 
    \frac{\Omega_1}{\pi^2\chi^2} +
    {\bar{a}_3}{\left(\frac{|\Omega_1|}{\pi^2\chi^2}\right)^{3/2}} + 
    {\bar{a}_4} {\left(\frac{\Omega_1}{\pi^2\chi^2}\right)^{2}}  +  
   \frac{\Omega_2}{\pi^4\chi^4} +
    {\cal{O}} {\left(\frac{|\Omega_1|^{5/2},\,|\Omega_1|^{1/2}\Omega_2} {\pi^5\chi^5} \right)}
      \right] \ . 
\end{equation}
 The leading  $f_2$ corrections appear  at the order $\Omega_2/(\pi \chi)^4$ while the corrections stemming from the quantum corrected geodesic $d_1$ appear at the order $(\Omega_1/(\pi^2 \chi^2))^{3/2}$ which is one order less in $1/\pi \chi$, consistently with this expansion. The coefficient $\bar{a}_3/4$ of $(\Omega_1/(\pi^2 \chi^2))^{3/2}$ is numerically evaluated  {in Appendix~\ref{App:NumericPosition} (see eq.~\eqref{numz+} for $\Omega_1>0$ and  eq.~\eqref{numz+1} for $\Omega_1<0$).}

\begin{figure}[!ht]
    \centering
    \includegraphics[width=0.8\textwidth]{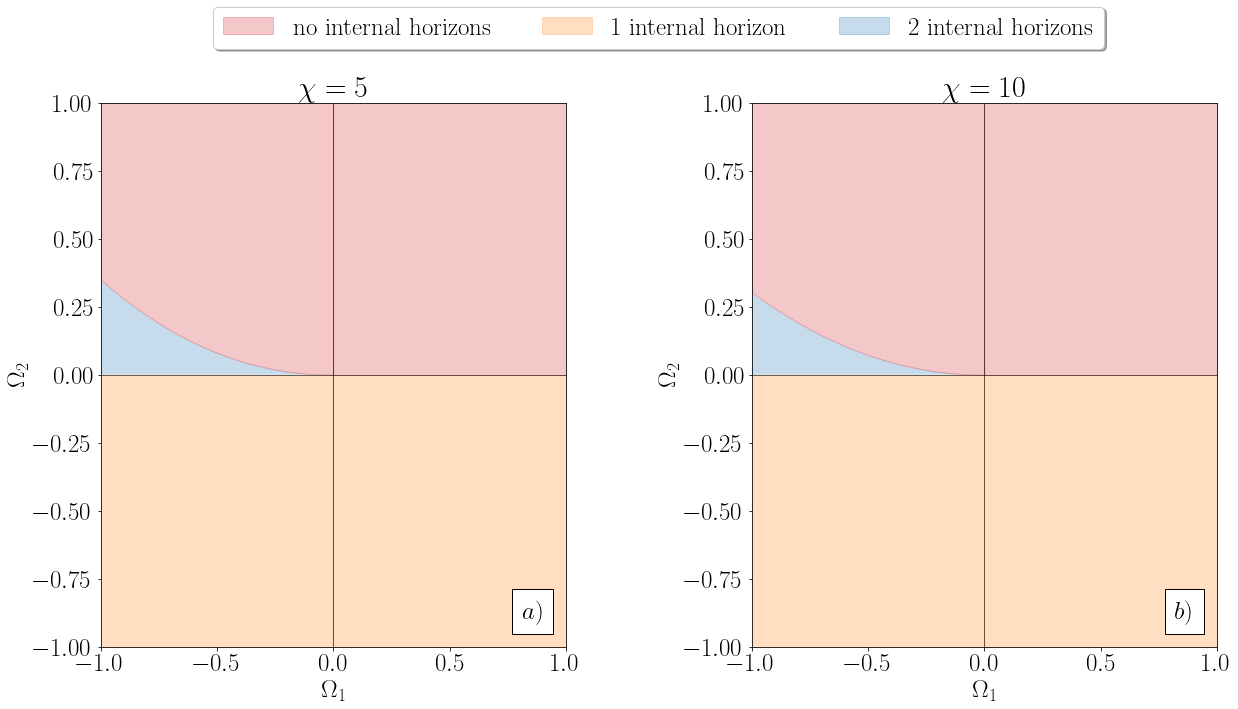}
    \caption{Phase diagram defining the regions in the $(\Omega_1,\Omega_2)$ plane with different numbers of internal horizons emerging from the quantum function $f_2$ for  $\chi=5$ (panel $a$) and $\chi=10$ (panel $b$).}
    \label{fig: o2 vs o1}
\end{figure}
The situation changes for the  {interior of the BH: as shown in Fig.~\ref{fig: fofrgeodesic2}, the structure of zeroes of the function $f_2$ depends crucially on the sign and magnitude of $\Omega_2$ (and $\Omega_1$). The BH geometry can therefore have zero, one or two internal horizons.}  {Concretely, the} phase diagram, representing the regions in the $(\Omega_1,\Omega_2)$ plane featuring different numbers of internal horizons stemming from the $f_2$ function, is shown in Fig.~\ref{fig: o2 vs o1} for  $\chi = 5$ in the left panel $a)$ and $\chi = 10$ in the right panel $b)$. One observes that the majority of the phase diagram features either none (upper pink region) or at most one (lower orange region) internal horizon.  The light blue region supports two internal horizons.

 \section{Thermodynamics}
 \label{Thermo}
So far we investigated the static properties of the BH in the effective quantum regime. We now move to determine its leading quantum thermodynamic properties.  

We start by computing the Hawking's \cite{Hawking:1975vcx} equilibrium temperature around the external horizon via its surface gravity parameter $\kappa$.  For the present BH metric the temperature is given in terms of the $f$ function, by
\begin{equation}\label{eq: hawking temp}
    T_\mathrm{H}=\frac{\kappa}{2\pi}=\frac{1}{4\pi}\left.\dv{f(r)}{r}\right|_{r=r_+}\, ,
\end{equation}
where $r_+$ is the radius of the external event horizon. Specializing this expression to the case in  \eqref{eq: schw Z quantum corrected} with $r_+ = \ell_{\rm P} z_+$, we  have
\begin{equation}\label{eq: schw hawking temperature}
    T_\mathrm{H}=\frac{1}{4 \pi \ell_\mathrm{P}}\left[\frac{2 \chi}{z_+^2}\left(1+\frac{\Omega_1}{d_0(z_+)^2} \right) + \frac{4 \chi}{z_+}\frac{\Omega_1}{d_0(z_+)^3} \left.\dv{d_0}{z}\right|_{z=z_+} \right]\, .
\end{equation}
The prefactor
\begin{equation}\label{eq: planck class temp}
    T_\mathrm{P}^{(0)}=\frac{1}{4\pi \ell_\mathrm{P}}\, ,
\end{equation}
can be naturally interpreted as the Hawking temperature for a classical Schwarzschild BH of Planck mass. From now on, we shall therefore work with the normalized ratio $T_\mathrm{H}/T_\mathrm{P}^{(0)}$.
 
\begin{figure}[!ht]
    \centering
    \includegraphics[width=0.9\textwidth]{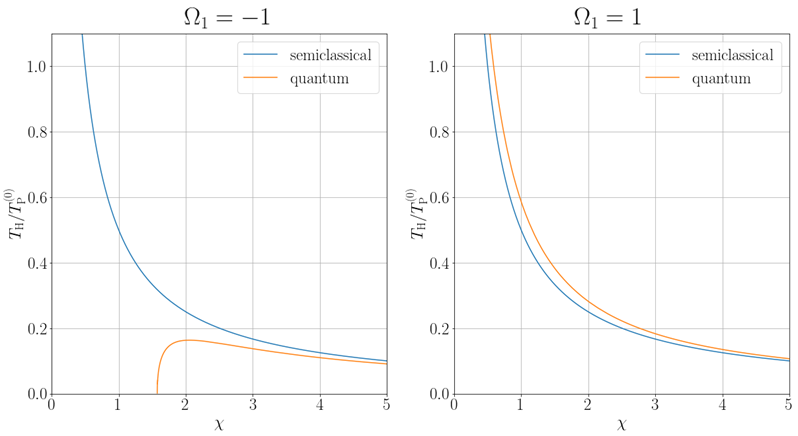}
    \caption{We plot the value of the Hawking temperature ratio $T_\mathrm{H}/T_\mathrm{P}^{(0)}$ as a function of $\chi$ in the case for $\Omega_1=-1$ (left) and $\Omega_1=1$ (right). It is noticeable that the temperature drops to zero at the extremal value of the mass while the \emph{(semi)}classical result continues evaporating.}
    \label{fig: schwarz temperature}
\end{figure}

As expected, for large enough values of $\chi$, the temperature  in  eq.~\eqref{eq: schw hawking temperature} tends to the \emph{semiclassical} {one, which is defined as}
\begin{equation}
   \frac{T_\mathrm{H}^{(0)}}{T_\mathrm{P}^{(0)}}=\frac{1}{2 \chi}\, .
\end{equation}
{This is shown in Fig.~\ref{fig: schwarz temperature}.} The consistent quantum expansion {of}  eq.~\eqref{eq: schw hawking temperature} reads:
\begin{equation}
    \frac{T_\mathrm{H}}{T_\mathrm{P}^{(0)}}=
    \begin{cases}\displaystyle
    \frac{1}{2\chi}\left[1-\frac{4}{\pi^2}\frac{\sqrt{|\Omega_1|}}{\chi}+\left(1-\frac{48}{\pi^2}\right)\frac{|\Omega_1|}{\pi^2\chi^2} +\mathcal{O}\left(\left( \frac{|\Omega_1|}{\pi^2\chi^2}\right)^{3/2} \right)\right] & \Omega_1 < 0 \ ,\\ \\\displaystyle
    \frac{1}{2\chi}\left[1+\frac{4}{\pi^2}\frac{\sqrt{|\Omega_1|}}{\chi}-\left(1+\frac{48}{\pi^2}\right)\frac{|\Omega_1|}{\pi^2\chi^2} +\mathcal{O}\left( \left( \frac{|\Omega_1|}{\pi^2\chi^2} \right)^{3/2} \right) \right] & \Omega_1 > 0 \ .
    \end{cases}\label{TemperatureBH}
\end{equation}  
At the quantum level the corrected  BH temperature decreases or increases, depending on whether the first order corrections to the metric are negative or positive. Additionally,  {in contrast to the}  classical case the decrease (increase) to the external horizon due to the quantum corrections leads to a decrease (increase) in the associated quantum temperature.

Although our counting scheme for quantum corrections limits the validity of our analysis to the results above it is interesting to discuss the small $\chi$ limit for negative $\Omega_1$.  We consider still the leading order correction to the metric (\emph{i.e.} we consider $n=1$ in eq.~(\ref{metricf})), which we approximate as (\ref{eq: Bonanno small r expansion}). In this case,  the quantum BH temperature achieves a maximum around $\chi \sim 2$ and rapidly decreases to zero for smaller $\chi$. At this point the internal and external horizons merge and the BH becomes extremal \eqref{S-extremal}.  
If this picture holds for the full quantum result this would suggest that the evaporation process leaves behind a stable remnant rather than observing a complete evaporation of the BH. {We hasten to add, however, that higher order corrections cannot consistently be neglected in this regime and may qualitatively change the picture.}

We {finally} turn our attention to the BH entropy {$S$} which 
 is obtained by integrating the first law of BH thermodynamics
\begin{equation}\label{eq: entropy diff}
    \dd M = T_\mathrm{H} \dd S \implies \dd S = \frac{\dd M}{ T_\mathrm{H}}=M_\mathrm{P}\frac{\dd \chi}{T_\mathrm{H}}\, .
\end{equation}
The temperature depends on the mass ratio $\chi$ once $\Omega_1$ is fixed and therefore we have
\begin{equation}\label{eq: entropy integral}
    S=M_\mathrm{P}\int \frac{\dd \chi}{T_\mathrm{H}(\chi)}\, .
\end{equation}
Inserting the mass expansion of the temperature in eq.~(\ref{TemperatureBH}), we have that the entropy assumes, up to a reference value, the following form
\begin{equation}
    S=\begin{cases}\displaystyle
    4\pi \chi^2 \left[ 1 + \frac{8}{\pi^2}\frac{\sqrt{|\Omega_1|}}{\chi} - 4 \left(1 - \frac{64}{\pi^2} \right)\frac{|\Omega_1|}{\pi^2\chi^2} \log \chi + \mathcal{O}\left(\left( \frac{|\Omega_1|}{\pi^2\chi^2} \right)^{3/2} \right)  \right] & \Omega_1<0 \ ,\\ \\ \displaystyle
    4\pi \chi^2 \left[ 1 - \frac{8}{\pi^2}\frac{\sqrt{|\Omega_1|}}{\chi} + 4 \left(1 + \frac{64}{\pi^2} \right)\frac{|\Omega_1|}{\pi^2\chi^2} \log \chi + \mathcal{O}\left( \left(\frac{|\Omega_1|}{\pi^2\chi^2} \right)^{3/2} \right)  \right] & \Omega_1>0 \ .
    \end{cases}
\end{equation}
 Thus, the entropy increases (decreases) for negative (positive) values of $\Omega_1$.  The leading quantum correction is linear in $\chi$ while only the subleading quantum correction receives corrections in the logarithm of the mass. In fact, the latter vanish at the Planck mass. Logarithmic corrections to the entropy of quantum BHs have also been found by using various other methods, see for example \cite{Fursaev:1994te,Kaul:2000kf,Carlip:2000nv}.

\section{Conclusions}
\label{Conclusions}
We investigated the quantum nature of black holes by employing an effective approach able to describe  quantum deviations from the classical results. We have studied in detail the Schwarzschild black hole, which is the simplest example in four dimensions, however, our approach can readily be extended to other types of geometries (which we plan to discuss in the future). Here we notably assumed that no other physical quantum gravity scale emerges besides the Planck one. 

Upon setting up the framework we determined the quantum corrections to the event horizon structure. To leading order in the quantum corrections, and depending on {their sign}, we showed that the black hole can either have a single horizon or develop a second internal one. We tested the robustness of our results by further considering both the {backreaction on the} quantum  proper distance   as well as the effects of higher order corrections. In this way we have demonstrated that the quantum corrections can be consistently organised into expansions dictated by inverse powers of the mass. We also note that these results do not hinge on a particular model of quantum gravity: indeed, different models    {simply} provide specific values of the coefficients $\Omega_n$  {for the quantum} corrections. For example, according to the {radiative} computations in \cite{Bjerrum-Bohr:2002fji}, one would have $\Omega_1 = -167/(30 \pi) \sim - 1.77$ and therefore the physics is the one stemming from a negative value of $\Omega_1$. Similarly one could match the coefficients to the prediction for the quantum metric stemming from different quantum gravity actions featuring, for example, higher curvature terms such as $f(R)$ theories \cite{Odintsov:1991nd,Elizalde:1994av,Akhundov:2006gh}. For a recent discussion about potential constraints on some of the $\Omega_n$ coefficients see \cite{Knorr:2022kqp}.  Alternatively, in the future, some of these coefficients could be experimentally determined.   To further test the robustness of our quantum framework we have shown how to take into account higher order corrections to the metric. We have even provided the explicit form of the next to next leading order quantum corrections to the external horizon. The results demonstrate the effectiveness and reliability of the expansion. We have also observed that the fate of the internal horizon, for $\Omega_1$ negative, is sensitive to higher order corrections.  We have consequently provided, to the next-to-next leading quantum order, the parameter space diagram illustrating the various scenarios for the internal horizon. Because of the nature of the expansion in \eqref{metricf} one can only address the ultimate fate of the singularity at the origin of the black hole within a specific model of quantum gravity  which would allow to resum the entire series nonperturbatively. We also provided the conformal diagrams for the quantum corrected black holes and determined the impact of the quantum corrections on the thermodynamic properties such as temperature and entropy.

Our approach  differs from the renormalization group improvement of a black hole space time in which the Newton constant is upgraded to an effective    running coupling  \cite{Bonanno:2000ep,Platania:2019kyx}. Within this latter framework it has been recently shown \cite{Held:2021vwd} that the renormalization group improvement at the level of the metric is coordinate-dependent while the approach  is applicable at the level of curvature invariants. Although in our framework we still work at the level of the black hole metric  our quantum modified metric depending only on physical quantities leads to coordinate independent observables.

If the quantum scale for gravity turns out to be lower than the Planck scale, our framework can take this into account by a simple rescaling of the dimensionless proper distance used in the definition of the quantum $f$ function of  \eqref{lapse-function}. A smaller quantum gravity scale can lead to sizable phenomenological effects. 

Overall, we have showed that quantum corrections to black hole physics can be organised in a powerful expansion in their mass that, already at the leading order, allows us to explore the quantum nature of black holes at distances that can be as close to the origin of the black hole, as few times the Planck length. The framework can be employed to investigate quantum corrections for other extended gravitational objects.

\subsection*{Acknowledgments}
We thank Matthias Blau, Alfio Bonanno, Cliff Burgess, Salvatore Capozziello, John Donoghue, Domenico Orlando, Roberto Percacci, Alessia Platania and Manuel Reichert  for comments and/or  enlightening discussions.

\appendix
\section{Quantum Corrected Horizons of the Schwarzschild Metric}
\label{Appendix}
\subsection{Range of Validity of the Quantum Corrected Horizon Position}\label{App:RadiusConvergence}
\begin{figure}[!ht]
    \centering
    \includegraphics[width=0.6\textwidth]{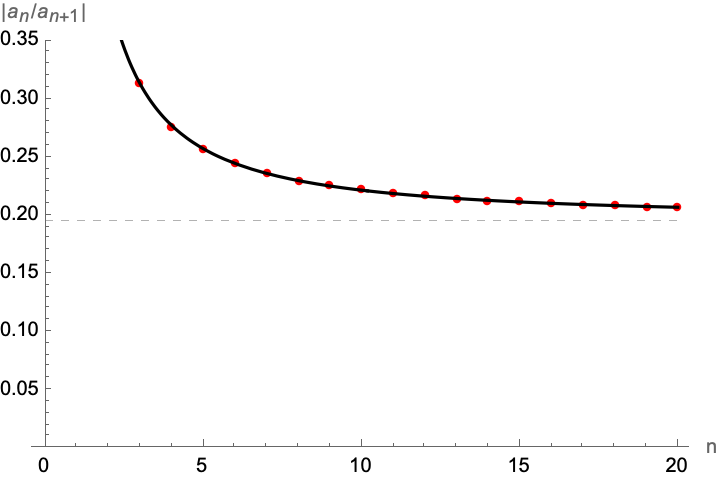}
    \caption{Absolute value of the quotient $|a_n/a_{n+1}|$ of the coefficients in eq.~(\ref{SeriesZ}). The black line interpolates the coefficients by a function of the form $\frac{a}{\alpha^b}+c$ for $a=0.475\pm 0.003$, $b=1.279\pm 0.008$ and $c=0.1968\pm 0.0004$.}
    \label{Fig:RadiusOpos}
\end{figure}

We provide a numerical estimate of the range of validity of the approximation (\ref{eq: external horzion positive omega}) for the position of the (external) horizon calculated as the zeroes of the $f$ function (\ref{eq: Bonanno small r expansion}). For simplicity (and since these computations only act as an order of magnitude estimate), we limit ourselves to the case $\Omega_1$. Specifically, we consider the zero (\ref{eq: external horzion positive omega}) as an entire series of the form
\begin{align}
    z_+=2\chi\sum_{n=0}^\infty a_n\,\alpha^{n/2}\,,&&\text{with} &&\begin{array}{l}a_0=1\,, \\ a_1=0\,, \\ a_2=1\,.\end{array}\label{SeriesZ}
\end{align}
An estimate for the range of validity of the result (\ref{eq: external horzion positive omega}) can be obtained by the radius of convergence of this series. Indeed, in  Fig.~\ref{Fig:RadiusOpos} we have plotted (norm) of the ratio $|a_n/a_{n+1}|$ which asymptotically approaches to the radius of convergence $r\sim 0.1968\pm 0.0004$. This suggests, that the result (\ref{eq: external horzion positive omega}) can be trusted for BH masses
\begin{align}
    \chi\geq \frac{\sqrt{\Omega_1}}{\pi\sqrt{r}}\sim 0.718\,\sqrt{\Omega_1}\,.
\end{align}

\subsection{Numerical Calculation of the Position of the Horizon(s)}\label{App:NumericPosition}
Here we perform a numerical analysis of the zeroes of the quantum self improved metric function \eqref{eq: schw Z quantum geodesic improved} with $d_1$ defined in \eqref{eq: quantum geodesic distance integrand}. We are in particular interested in the order in the parameter $\alpha$ (defined in eq.~\eqref{eq: external horzion positive omega}) at which this modification changes the position of the horizon(s) of the black hole compared to $f_1$ defined in (\ref{eq: schw Z quantum corrected}). To this end, we distinguish the cases $\Omega_1>0$ and $\Omega_1<0$
\begin{itemize}
\item $\Omega_1>0$: In this case, the zeroes of $\bar{f}_1$ as a function of $\alpha$ (for $\chi=2$) are plotted in Figure~\ref{Fig:QcorrectedOpos} for the choice $\chi=2$. The solid black line is approximated by
\begin{figure}[!ht]
    \centering
    \includegraphics[width=0.6\textwidth]{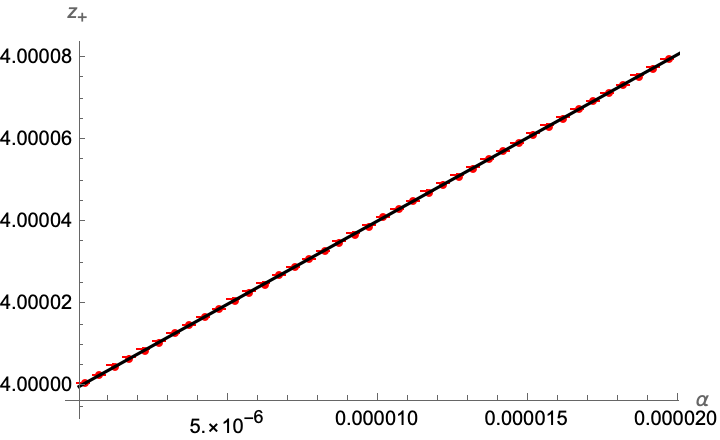}
    \caption{Numerical position of the position of the horizon $z_+$ as a function of $\alpha$ for $\chi=2$ and $\Omega_1>0$.}
    \label{Fig:QcorrectedOpos}
\end{figure}
\begin{align}
\label{numz+}
&z_+=4(\bar{a}_1+\bar{a}_2\,\alpha+\bar{a}_3\,\alpha^{3/2})\,.&&\text{with}&&\begin{array}{l}4\bar{a}_1=4\pm 2.2*10^{-9}\,, \\ 4\bar{a}_2=4.0036\pm 0.007\,, \\ 4\bar{a}_3=7.853\pm2.523\,.\end{array}
\end{align}
The numerical results for $\bar{a}_1$ and $\bar{a}_2$ are compatible with the coefficients obtained for the zeroes of $f_1$, while the coefficient $\bar{a}_3$ is not (namely $a_1=1$, $a_2=1$ and $a_3=-\frac{2}{\pi}\sim -0.637$). This suggests, that the corrections to the position of the horizon $z_+$ that stem from using the improved metric function $\bar{f}_1$ are of order $\mathcal{O}(\alpha^{3/2})$.
\item $\Omega_1<0$:  In this case, the two zeroes of $\bar{f}_1$ as a function of $\alpha$ (for $\chi=2$) are plotted in Figure~\ref{Fig:QcorrectedOneg} for $\chi=2$. The solid black line is approximated by
\begin{figure}[!ht]
    \centering
    \includegraphics[width=0.4\textwidth]{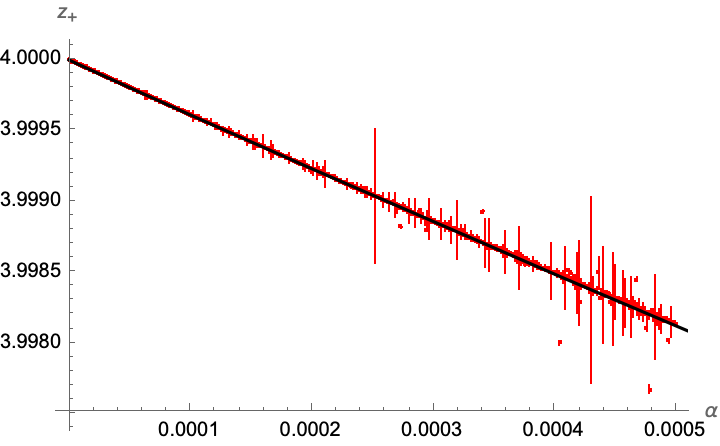}\hspace{1cm} \includegraphics[width=0.4\textwidth]{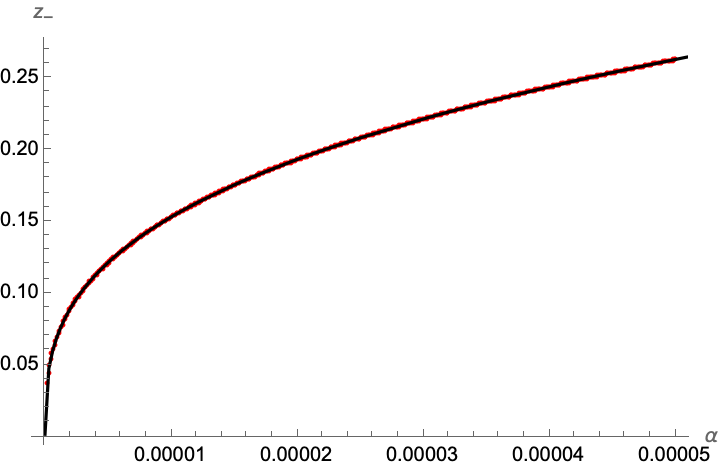}
    \caption{Numerical position of the position of the horizon $z_+$ (left panel) and $z_-$ (right panel) as a function of $\alpha$ for $\chi=2$ and $\Omega_1<0$.}
    \label{Fig:QcorrectedOneg}
\end{figure}
\begin{align}
\label{numz+1}
&z_+=4(\bar{a}_1+\bar{a}_2\,\alpha+\bar{a}_3\,\alpha^{3/2})\,,&&\text{with}&&\begin{array}{l}4\bar{a}_1=4\pm 4.8*10^{-7}\,, \\ 4\bar{a}_2=-3.979\pm 0.045\,, \\ 4\bar{a}_3=9.947\pm2.685\,,\end{array}\\
&z_-=\bar{b}_1\,\alpha^{\frac{1}{3}}+\bar{b}_2\,\alpha^{\frac{2}{3}}+\bar{b}_3\,\alpha+\bar{b}_4\,\alpha^{\frac{4}{3}}\,,&&\text{with}&&\begin{array}{l}\bar{b}_1=7.0827\pm 0.00005\,, \\ \bar{b}_2=1.6724\pm 0.0114\,. \\ \bar{b}_3=2.26\pm0.60\,,\\ \bar{b}_4=4.45\pm 9.08\,.\end{array}
\end{align}
For the position of the outer horizon $z_+$, the situation is similar to the case $\Omega_1>0$: the coefficients $\bar{a}_{1,2}$ are numerically compatible with the values $a_{1,2}$ coming from $f_1$, while $\bar{a}_3$ is not. This suggests, that the corrections to the position of the horizon $z_+$ that stem from using the improved metric function $\bar{f}_1$ are of order $\mathcal{O}(\alpha^{3/2})$. For $z_-$ the coefficients $\bar{b}_{1,2,3}$ are numerically compatible with the values obtained from $f_1$. In view of the large numerical uncertainty, this suggests that the corrections to the position of the inner horizon appear at an order larger than $\mathcal{O}(\alpha)$.
\end{itemize}
 
\subsection{Second Order Quantum Corrections}\label{App:SecondOrderCorr}
Before treating the second order quantum corrections, we first provide a relation between the proper distances $d_1$ and $d_0$: we start from the definition of $d_1$ in (\ref{eq: quantum geodesic distance integrand}), however, instead of a function of $z$ we consider it as a function of $d_0$ defined in (\ref{eq: schwarz geodesic distance}).\footnote{This is possible since $d_0$ is a monotonic function in $z$.} We then find for the first derivative
\begin{align}
\frac{\dd d_1}{\dd z}=\frac{\dd d_1}{\dd d_0}\,\frac{\dd d_0}{\dd z}=\frac{1}{\sqrt{\left|1-\frac{2\chi}{z}\left(1+\frac{\Omega_1}{d_0^2}\right)\right|}}\,.
\end{align}
Using (\ref{eq: schwarz geodesic distance integrand}) we therefore find
\begin{align}
\frac{\dd d_1}{\dd d_0}=\sqrt{\frac{\left|1-\frac{2\chi}{z}\right|}{\left|1-\frac{2\chi}{z}\left(1+\frac{\Omega_1}{d_0^2}\right)\right|}}\label{DiffeqDistance}
\end{align}
where implicitly $z$ is understood as a function of $d_0$. We can consider two limits of this equation
\begin{itemize}
\item $\chi \ll z$ (or equivalently $\chi\ll d_0$): in this case, the equation (\ref{DiffeqDistance}) becomes
\begin{align}
\frac{\dd d_1}{\dd d_0}=1\,,
\end{align}
which has as solution $d_1\sim d_0$, \emph{i.e.} for distances far away from the BH the two distances become equivalent.
\item $\chi \gg z$ (or equivalently $\chi\gg d_0$): in this case, the equation (\ref{DiffeqDistance}) becomes
\begin{align}
\frac{\dd d_1}{\dd d_0}=\sqrt{\frac{1}{\left|1+\frac{\Omega_1}{d_0^2}\right|}}\,,
\end{align}
which is a differential equation for $d_1$ and can be integrated up in a direct fashion\footnote{Requiring $d_1$ to be $u$ independent implies also independence of $\Omega_1$ on $u$.}
\begin{align}
d_1=\left\{\begin{array}{lcl}d_0\sqrt{1+\frac{\Omega_1}{d_0^2}}-\sqrt{\Omega_1} & \text{if} & \Omega_1> 0 \,,\\ \sqrt{-\Omega_1}-d_0\sqrt{-\frac{\Omega_1}{d_0^2}-1} & \text{if} & \Omega_1<0 \text{ and } d_0<\sqrt{-\Omega_1})\,,
\\ \sqrt{-\Omega_1}+d_0\sqrt{1+\frac{\Omega_1}{d_0^2}} & \text{if} & \Omega_1<0 \text{ and } d_0>\sqrt{-\Omega_1})\,.\end{array}\right.\label{d1sol}
\end{align}
Here the integration constants have been chosen in such a way that $d_1$ is a continuous function of $d_0$ and  $\lim_{d_0\to 0}d_1=0$. Graphically, the solutions are shown in Figure~\ref{Fig:d1Fund0}. For small values of $d_0$ we find

\begin{figure}[!ht]
    \centering
    \includegraphics[width=0.4\textwidth]{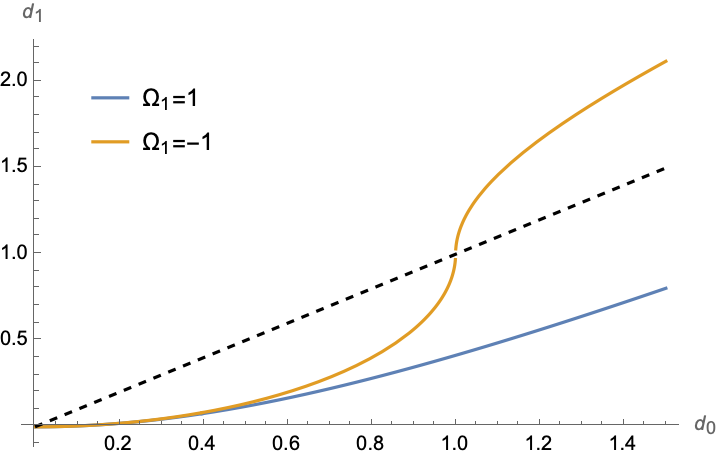}\hspace{0.5cm} \includegraphics[width=0.4\textwidth]{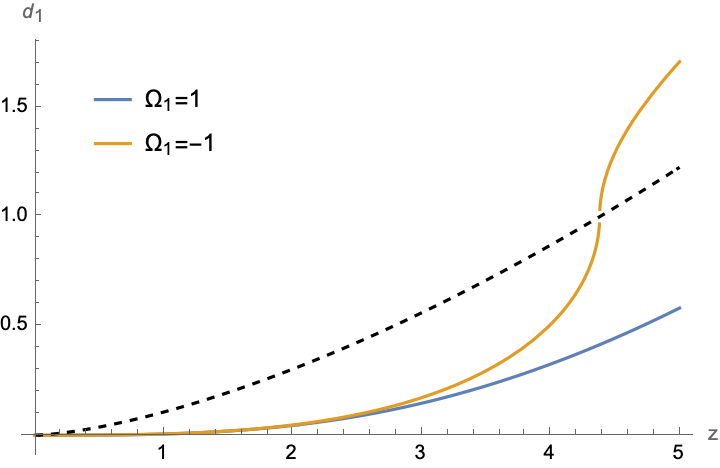}
    \caption{Distance $d_1$ in eq.~(\ref{d1sol}) as a function of $d_0$ (left panel) and of $z$ assuming that $\chi=20$ (right panel). The dashed black line corresponds to $d_0$ for comparison.}
    \label{Fig:d1Fund0}
\end{figure}

\noindent
\begin{align}
d_1\sim \frac{d_0^2}{2\sqrt{|\Omega_1|}}+\mathcal{O}(d_0^4)\,.
\end{align}
\end{itemize}

\noindent
We can now turn to the second order quantum corrections. To this end, we generalise the differential equation (\ref{DiffeqDistance}) for $d_2$ (as a function of $d_1$)
\begin{align}
\frac{\dd d_2}{\dd d_1}=\sqrt{\frac{\left|1-\frac{2\chi}{z}\left(1+\frac{\Omega_1}{d_0^2}\right)\right|}{\left|1-\frac{2\chi}{z}\left(1+\frac{\Omega_1}{d_1^2}+\frac{\Omega_2}{d_1^4}\right)\right|}}\,.\label{Diffeqd2}
\end{align}
As before, we can consider two limits of this equation
\begin{itemize}
\item $\chi \ll z$ (or equivalently $\chi\ll d_1$): in this case, the equation (\ref{Diffeqd2}) becomes
\begin{align}
\frac{\dd d_2}{\dd d_1}=1\,,
\end{align}
which has as solution $d_2\sim d_1$, \emph{i.e.} for distances far away from the BH the two distances become equivalent.
\item $\chi \gg z$ (or equivalently $\chi\gg d_1$): in this case, the equation (\ref{Diffeqd2}) becomes
\begin{align}
\frac{\dd d_2}{\dd d_1}=\sqrt{\frac{\left|1+\frac{\Omega_1}{d_0^2}\right|}{\left|1+\frac{\Omega_1}{d_1^2}+\frac{\Omega_2}{d_1^4}\right|}}\,.\label{Diffeqd2b}
\end{align}
which together with (\ref{d1sol}) is a differential equation for $d_2$ and can in principle be integrated as a function for $d_1$. Since this is, however, technically difficult, we focus on expanding $d_2$ around $d_1=0$ and for simplicity focus on the case $\Omega_{1,2}>0$\footnote{Other cases can be analysed in a similar fashion}:
\begin{align}
\frac{\dd d_2}{\dd d_1}=\sqrt{\frac{1+\frac{\Omega_1}{d_1(d_1+2\sqrt{\Omega_1})}}{1+\frac{\Omega_1}{d_1^2}+\frac{\Omega_2}{d_1^4}}}\,.
\end{align}
Expanding for small $d_1$ (or equivalently small $d_0$), we find
\begin{align}
d_2\sim \frac{\sqrt{2}}{5}\,\frac{\Omega_1^{1/4}}{\sqrt{\Omega_2}}\,d_1^{5/2}+\mathcal{O}(d_1^{7/2})\sim \frac{d_0^5}{20\,\Omega_1\,\sqrt{\Omega_2}}+\mathcal{O}(d_0^7)\,,
\end{align}
which is indeed compatible with (\ref{Scalingd1d2}).
\end{itemize}

\printbibliography
\end{document}